\documentclass[12pt]{iopart}

\usepackage{iopams}
\usepackage{cite}
\usepackage{graphicx}

\begin{document}

\title[A Dynamical Mechanism for the Big Bang and Non-Regularizability for $w=1$]{A Dynamical Mechanism for the Big Bang and Non-Regularizability for $w=1$}

\author {Edward Belbruno}
\address{ Department of Astrophysical Sciences,}
\address{Princeton University, Princeton, NJ 08544, USA}
\eads{\mailto{belbruno@princeton.edu}}

\begin{abstract}
 We consider a contracting universe and its transition to expansion through the big bang singularity with a time varying equation of state  $w$, where $w$ approaches $1$ as the universe contracts to the big bang. We show that this singularity is non-regularizable. That is, there is no unique extension of the physical quantities after the transition, but rather infinitely many.  This is entirely different from the case of $w > 1$ studied in \cite{Xue:2014},  where $w$ approaches a constant value $w_c > 1$ as the universe contracts. In that case a continuous transition through the big bang to yield a unique extension was possible only for a discrete set of $w_c$ satisfying coprime conditions.    We also show that there exists another time variable, $N$, at the big bang singularity itself, at $t=0$,  where  $w$, varies as a function of $N$. This defines an {\em extended big bang state}.  Within it, $H$ is infinity. In the extended state, $w$ varies from a universe dominated by the cosmological constant to $1$. After $w$ reaches $1$ then the big bang occurs and time $t >0$ resumes. This gives a dynamical mechanism for the big bang that is mathematically complete as a function of $N$. Dynamical systems methods are used with classical modeling.
  \end{abstract}


\submitto{\CQG}

\maketitle

\section{Introduction} \label{sec:intro} 

The big bang is a singularity in space and time where the universe begins and starts to expand. The inflationary theories  describe a very fast expansion from that singular state. We study this state by first considering a contracting universe that contracts to the big bang and then try and understand under what conditions it can make a transition to an expanding universe, as is done in bouncing cosmologies. We do this from a dynamical systems perspective. That is, given the dynamics of the physical quantities, what are the conditions under which the solutions of the physical quantities can be continuously extended through the singularity?  In \cite{Xue:2014} this was studied for the case where the equation of state for the dominant energy component is time varying, where the energy component is modeled as a scalar field. The basic assumption in that study was that as the universe contracts to the big bang at $t=0$, as $ t \rightarrow 0$ for $ t < 0$, then $ w \rightarrow w_c > 1$.    The main result was that a continuous extension of the solutions of the physical quantities can be obtained to $t \geq 0$ provided that $w_c$ satisfies coprime conditions, which is a discrete set of values. In fact, each extension is unique. 

In this paper, we consider the special case where $w \rightarrow 1$ as $t \rightarrow 0$. In this case, we mathematically show that the nature of big bang singularity changes, and unique extensions of the physical quantities are not possible. Instead, one has infinitely many possible extensions, all equally likely.  This is an interesting result, and yields infinitely many possibilities for the expanding universe. The nonuniqueness of extensions implies that the the big bang cannot be regularized which is a main result of this paper.

By studying this case, we can also obtain an interesting mechanism for the big bang which extends the big bang singularity and gives it a dynamical structure. We show mathematically how the big bang singularity itself at $t=0$  can be extended  using another time variable, $N$, where $w$ varies. It is mathematically complete as a function of N, since $-\infty < N < \infty$.  As  $N$ varies in this range, with $t =0$,  $-1 <  w <  1$, respectively, and the scale factor, $a =a(N)$ varies from $\infty$ to $0$, respectively,  but $Q = 1/H = 0$, where $H$ is the Hubble variable. That is, each point of the universe has infinite energy density with $a > 0$.  We call this the {\it extended big bang state}.  At the boundary of this state, when $N = \infty$,  $w=1$ and $a = 0$. The big bang occurs and the universe expands for $t > 0$.

It is noted that in inflationary cosmology, the space-time singularity is shown to be incomplete in the past \cite{Borde:2001nh}. This implies that there is a singularity in the finite past. However, this does not imply that  there cannot be an induced time within the big bang singularity itself, which yields a mathematically complete structure.

A mechanism for the big bang is described as follows:  The big bang singularity is described by the extended big bang state at $t=0$. It  begins when $w=-1$ for the cosmological constant when $N = -\infty$, $a= \infty$.   As $N$ increases to $+\infty$, $w$ increases to $1$. As this variation occurs, $Q = 0$, and $a>0$ varies from $a = \infty$ to $a=0$ when $w=1$ is reached, then $t >0$,  $a >0$ with $Q > 0$. Then $a, Q, w$ increase. This is the observed big bang itself.   Within the extended big bang state $t=0$, so that the variation of $a, w$ as a function of $N$ is that of their values at $t=0$: $a= a_0 \equiv a|_{t=0} = 0, w=w_0\equiv w|_{t=0} = 1$.  This represents a covering of $a_0, w_0$, parameterized by $N$. (A covering of $a_0$  is the set, $A = \{ 0 < a < \infty \}$ and a continuous map of $A$ onto the point $a_0$, and a covering of $w_0$ is the set, $B = \{ -1 < w < 1\}$ and a continuous map of $B$ onto the point $w_0$. See Section \ref{sec:2})
\medskip

The mathematical approach we take here is analogous to that taken initially in  \cite{Belbruno:2013}, for constant $w$, and then much more generally in \cite{Xue:2014}  when $w$ is time varying, where the big bang singularity is analyzed using dynamical systems methods. In particular, a regularization transformation is applied to the dynamical system consisting of the Friedmann equations for the Hubble variable $H$ and the differential equation for $w$. The purpose of this transformation is to obtain a new set of coordinates, including time, where the flow of the transformed differential equations in a neighborhood of the big bang is well defined.  This is done in order to understand whether a solution $a(t), w(t)$, defined for $t < 0$, where $a \rightarrow 0, w \rightarrow w_c$ as $ t \rightarrow 0$, could be shown to be well defined at $t=0$ and also give rise to a unique extension, $a(t), w(t)$, for $t>0$.  This is called a branch regularization. As is  rigorously shown in \cite{Xue:2014} a branch regularization exists if and only if $w_c > 1$ and $w_c$ satisfies special coprime number conditions (a set of measure zero). Therefore, with the exception of this set of measure zero, there is no branch extension. The proof of this utilizes the stable manifold theorem \cite{Belbruno:2004, Guckenheimer:2002}, which requires that the eigenvalues of the constant $2x2$ matrix,  representing the leading order linear terms of the differential equations are non-zero and real.  
\footnote{The existence of the branch regularization for the relative prime number set of $w_c$, has relevance to the bouncing cosmology theory, where there is a contraction phase before the expansion phase.  Under general conditions as considered in the singularity theorem \cite{Hawking:1969sw}, the cosmic contraction would end in a spacetime singularity. In that scenario, the big bang is both the end of the contraction phase and the beginning of the expansion phase.}

It turns out that no conclusions can be made on extending solutions through the big bang using the methods in \cite{Xue:2014} when $w_c = 1$ since in that case one of the eigenvalues of the $2x2$ matrix vanishes.  This then yields a degenerate situation where one of the leading order terms of the the differential equations is no longer linear, but quadratic. This degeneracy yields nonuniqueness of solutions at the singularity. As is shown in this paper this degeneracy can be overcome in a straight forward manner by appealing to a special mathematical theorem.  In this case, for each solution $a(t), w(t)$, $t < 0$,  
$a \rightarrow 0, w \rightarrow w_c = 1$, as $ t \rightarrow 0$, there exists {\em infinitely} many possible extensions $a(t), w(t)$ for $t > 0$.  Thus, branch regularization is not possible.

The starting point for the big bang can be viewed as first starting from the contracting universe, where each solution first enters the extended big bang state, as $N$ varies from $-\infty$ to $\infty$, and $a > 0$ varies from $0$ to $\infty$, respectively,  $w$ varies from $1$ to $-1$, respectively. The direction of flow in the big bang state then reverses, and the extended big bang leads to the big bang as described previously.  

On the other hand, the beginning of the big bang can be viewed as starting from the extended big bang state for $t=0$,  with $N = -\infty$, $w=-1$, $a= \infty$ and then to the big bang itself with  $a=0$, $N = \infty$, $w=1$. This leads to a possible interpretation of this mechanism:  One can imagine that after the big bang, as time $t$ increases, the universe expands over trillions of years where all the matter and energy have dissipated and all that is left is pure space. At this epoch, $a = \infty$ and $w = -1$.  But then at that moment,  every point is equivalent to any other point and the universe collapses to a single point where $a=0$. Time $t$ is set to $0$. At $t=0$,  $N$ varies from $-\infty$ to $\infty$, where $a$ varies from $\infty$ to $0$, respectively. A scenario like this was  described in a short story by I. Asimov \cite{Asimov:1956}. This is also discussed in Section \ref{subsec:differentmechanisms}.

The model we assume in this paper is for a flat, homogeneous, isotropic universe with the standard Friedmann equations and a time varying equation of state. Also, the scalar field is modeled with an exponential potential, which is widely used in inflationary theories \cite{Burd:1988ss}.

This paper is organized as follows. Section~\ref{sec:1} presents the system of dynamical equations for a set of physical quantities that we consider. Section~\ref{sec:2} gives the main results on non-regularizability of the big bang and possible mechanisms for the big bang.  Physical interpretations are given.  This is generalized in Section~\ref{sec:3} to additional energy components. The results are summarized in Section~\ref{sec:4}.

\section{Dynamical system} \label{sec:1} 

In this section we derive the differential equations for a set of variables that describe the evolution of the universe. The Friedmann equations lead to a differential equation for the Hubble parameter $H$, or its reciprocal, $Q \equiv 1/H$. The equation of motion for the scalar field $\phi$ determines its time varying equation of state $w$. With additional energy components besides the scalar field, we introduce the relative energy density $\Omega_m$ for each component and derive the differential equations for them.

Consider a homogeneous, flat, and isotropic universe with the metric
\begin{equation}
ds^2 = - dt^2 + a(t)^2 |d{\bf x}|^2 ,
\end{equation}
where $t$ is the proper time and $\mathbf{x} = (x^1, x^2, x^3)$ are the spatial coordinates. Here $a(t)$ is the scale factor of the universe, and the Hubble parameter $H$ is given by $H \equiv \dot{a} / a$, where the dot $\dot{}$ denotes the derivative with respect to time $t$. $H$ is negative during cosmic contraction, and positive after the universe transitions to expansion. The big bang (or ``big bounce'') is at $a = 0$, which is chosen to correspond to the time $t = 0$.

Assume that the scalar field $\phi$ has the Lagrangian
\begin{equation}
\mathcal{L} = \sqrt{-g} \Big[ -\case{1}{2} (\partial \phi)^2 - V(\phi) \Big] ,
\end{equation}
where the potential $V(\phi)$ is an exponential function $V(\phi) = V_0 \, e^{- c \, \phi}$; it is assumed that $V_0 < 0$ and $c  = $constant, explained below. In the homogeneous case, the energy density and pressure of the scalar field are
\begin{equation}
\rho_\phi = \case{1}{2} \, \dot{\phi}^2 + V(\phi), \quad p_\phi = \case{1}{2} \, \dot{\phi}^2 - V(\phi).
\label{eq:rhoP}
\end{equation}
The equation of state parameter $w$ is given by
\begin{equation}
w = {p_\phi} / {\rho_\phi} .
\label{eq:w}
\end{equation}
The equation of motion for the $\phi$ field can be written as (see \cite{Xue:2014})
\begin{equation}
\dot{\rho}_\phi + 3 H (1 + w) \rho_\phi = 0 .
\label{eq:rhoDE}
\end{equation}

Consider first the case in which the scalar field $\phi$ is the only energy component in the contracting universe. 
Then the Hubble parameter $H$ obeys the Friedmann equations,
\numparts
\begin{eqnarray}
H^2 = \case{1}{3} \, \rho_\phi , \label{eq:FriedmannEqus1} \\*[4pt]
\dot{H} = -\case{1}{2} (\rho_\phi + p_\phi) . \label{eq:FriedmannEqus2}
\end{eqnarray}
\endnumparts
Using equation~(\ref{eq:w}), the Friedmann equations yield a differential equation for $H$,
\begin{equation}
\dot{H} = - \case{3}{2} (1 + w) H^2 .
\label{eq:HDiffEqu}
\end{equation}
Since $H$ is negative during contraction, for $w > -1$ (as required by the null energy condition), $H \to -\infty$ at the big bang.

It is also shown in \cite{Xue:2014} that
\begin{equation}
\dot{w} = \frac{3H \sqrt{1 + w}}{\sqrt{1 + w_c} + \sqrt{1 + w}} \, (w - 1) (w - w_c) ,
\label{eq:wdot}
\end{equation}
where 
\begin{equation}
w_c \equiv \frac{c^2}{3} - 1 ,
\end{equation}
and  $c = -V_{\phi} / V $, $V_{\phi} \equiv \partial{V}/\partial{\phi}$.
\medskip

It can be seen that, for $c > \sqrt{6}$ and hence $w_c > 1$, $w = w_c$ is a fixed point attractor in a contracting universe where $H < 0$.  This restriction is assumed in \cite{Xue:2014}.  Note that the attractor solution is described by $\dot{\phi}= 2 / (c \, t)$, where $t < 0$ during contraction. Hence, as $t \rightarrow 0$,  $\phi \rightarrow -\infty. $ 

In this paper, we consider the limiting case $w_c=1$, or equivalently, $c = \sqrt{6}$.    In this case,  the fixed point solution $w_c=1$ is no longer an attractor.

If there exist other energy components in the universe in addition to the scalar field $\phi$, then equations~(\ref{eq:HDiffEqu}) and (\ref{eq:wdot}) need to be modified. As is described in \cite{Xue:2014}, consider energy components with constant equations of state $w_m$, where $w_m = 0$, $\frac{1}{3}$, $-1$, $-\frac{1}{3}$, or $1$, if the additional energy components represent matter, radiation, cosmological constant, spatial curvature, or anisotropy, respectively. The energy density $\rho_m$ of each component obeys an equation similar to (\ref{eq:rhoDE}),
\begin{equation}
\dot{\rho}_m + 3 H (1 + w_m) \rho_m = 0 .
\label{eq:rhomDE}
\end{equation}
To describe the extra degrees of freedom, introduce the density parameters $\Omega_m$ defined by
\begin{equation}
\Omega_m \equiv \rho_m / \rho_{\rm tot} , \quad \rho_{\rm tot} = \rho_\phi + \sum_m \rho_m ,
\end{equation}
which represent the fractional density of each energy component. The Friedmann equations~(\ref{eq:FriedmannEqus1}, \ref{eq:FriedmannEqus2}) are modified as
\numparts
\begin{eqnarray}
H^2 = \case{1}{3} \, \rho_{\rm tot} = \case{1}{3} \sum_i \rho_i , \label{eq:FriedmannEqusOne} \\*
\dot{H} = - \case{1}{2} (\rho_{\rm tot} + p_{\rm tot}) = -\case{1}{2} \sum_i \rho_i (1 + w_i) , \label{eq:FriedmannEqusTwo}
\end{eqnarray}
\endnumparts
where in these two equations the summation is over all energy components including the scalar field $\phi$.

For illustration, consider just one additional component besides the scalar field. Let the equation of state and the density parameter of this component be denoted by $w_1$ and $\Omega_1$. The Friedmann equations yield the differential equation
\begin{equation}
\dot{H} = - \case{3}{2} \big[ (1 + w) - (w - w_1) \Omega_1 \big] H^2 .
\label{eq:HDiffEquTwo}
\end{equation}
Similarly, equation~(\ref{eq:wdot}) for the equation of state $w$ of the scalar field $\phi$ is modified as
\begin{equation}
\dot{w} = \case{3 \sqrt{w + 1} \, H}{\sqrt{w + 1} + \sqrt{(w_c + 1) (1 - \Omega_1)}} \, (w - 1) \big( w-w_c + (1 + w_c) \Omega_1 \big).
\label{eq:wdot2Two}
\end{equation}
Using (\ref{eq:rhoDE}) and (\ref{eq:rhomDE}), one obtains an additional equation for $\Omega_1$,
\begin{equation}
\dot{\Omega}_1 = 3 H (w - w_1) \Omega_1 (1 - \Omega_1) .
\label{eq:dotOmega}
\end{equation}
We assume that $w_1 < 1 \leq w_c$. Ths is true for any $w_m \in \{ 0, \frac{1}{3}, -1, -\frac{1}{3}, 1 \}$ except $w_1 = 1$ which is excluded. Equations~(\ref{eq:HDiffEquTwo}, \ref{eq:wdot2Two}, \ref{eq:dotOmega}) have a singularity at the big bang when $H \to -\infty$. Therefore the solutions of $H(t)$, $w(t)$, and $\Omega_1(t)$ need to be regularized in order to extend through the singularity.

We introduce the variables $Q \equiv 1/H$ and $W \equiv w - w_c$, $w_c=1$,  which both go to $0$ at the singularity. Then equations~(\ref{eq:HDiffEquTwo}, \ref{eq:wdot2Two}, \ref{eq:dotOmega}) can be written in terms of these new variables as follows.

\medskip

\noindent
\textbf{Summary~1.} \; For a homogeneous universe filled with the scalar field $\phi$ and an additional energy component with a constant equation of state $w_1$, the variables $Q$, $W$, and $\Omega_1$ satisfy a system of differential equations
\numparts
\begin{eqnarray}
\hspace{-0.5in} \dot{Q} = \case{3}{2} \Big[ (W + w_c + 1) - (W + w_c - w_1) \Omega_1 \Big] , \label{eq:DiffEqusMoreGeneralH} \\*[4pt]
\hspace{-0.5in} \dot{W} = \case{3 \sqrt{W + w_c + 1} \, / Q}{\sqrt{W + w_c + 1} + \sqrt{(w_c + 1) (1 - \Omega_1)}} \, (W + w_c - 1) \big( W + (1 + w_c) \Omega_1 \big) , \label{eq:DiffEqusMoreGeneralW} \\*[4pt]
\hspace{-0.5in} \dot{\Omega}_1 = \case{3 (W + w_c - w_1)}{Q} \, \Omega_1 (1 - \Omega_1) . \label{eq:DiffEqusMoreGeneralOmega}
\end{eqnarray}
\endnumparts

\medskip

\medskip

The key case that we consider first just has one component, the scalar field $\phi$.  We will also consider the generalization to one additional energy component, $\Omega_1$,  with a constant equation of state $w_1$, as presented in Summary~1. For clarity, we denote the density parameter $\Omega_1$ of the additional component simply by $\Omega$.

\section{Main Results} \label{sec:2}

The main result on the mechanism for the big bang is given.  It results from examining the problem of extending solutions through the big bang. 
We consider the solutions $a(t)$, $Q(t)$, $W(t)$, $\Omega(t)$ for $t < 0$, which tend to $a = 0$, $Q = 0$, $W = 0$, $\Omega = 0$, when $t \to 0^-$ (i.e. $t \rightarrow 0$ and $t < 0$), where $W = w-1$.    We determine necessary and sufficient conditions for these solutions to have a well defined unique extension to $t \geq 0$. The main case we consider first is for just a scalar field, where there are no other energy components, so that $\Omega(t) =0$.  

In the remainder of this paper, we will use the results from \cite{Xue:2014} and make reference to this paper frequently, as the basic equations and definitions are given there.

The first concept we need is that of branch regularization, defined in \cite{Xue:2014, Belbruno:2013}. This was defined generally in the Introduction, but more precisely, we would like to prove under what conditions, if any, that a solution ${\bf{X}_1}(t) = (Q(t), W(t))$ as $t \to 0^-$ can be uniquely extended to a solution ${\bf{X}_2}(t)$ for $t > 0$, and where ${\bf{X}_1}(0) ={ \bf{X}_2}(0) = (0,0)$

Having defined the terms, let us state the main result of the paper. Let $t_0 < 0$ be a certain time in the contracting phase of the universe prior to the big bang at $t = 0$. Consider the system of differential equations given by (\ref{eq:DiffEqusMoreGeneralH}, \ref{eq:DiffEqusMoreGeneralW}) for $Q(t), W(t)$ with initial conditions $Q(t_0) < 0$, $W(t_0)  \geq 0$. We have the following theorem.

\medskip

\noindent
\textbf{Result~1 (Non-Uniqueness)} \; The solution $(Q(t), W(t))$ of the dynamical system (\ref{eq:DiffEqusMoreGeneralH},\ref{eq:DiffEqusMoreGeneralW}) (with $\Omega = 0, w_c=1$) as well as $a(t)$, are not  branch regularizable at the singularity $t = 0$ , i.e.  there is no unique extension to $t > 0$, with $Q > 0$, for each solution  $(Q(t), W(t))$ with initial time  $t_0 <0$, where  $Q(t_0) < 0$, $W(t_0)  \geq 0$.  There are infinitely many possible extensions.     
\medskip\medskip

Consideration of an arbitrary solution $(Q(t), W(t))$ with initial time  $t_0 <0$, where  $Q(t_0) < 0$, $W(t_0)  \geq 0$, leads to 
\medskip\medskip

\noindent
\textbf{Result~2 (Extended Big Bang State)} \; The big bang singularity at $t=0$, defined by $a=0$, $W=0$, can be extended to the set, $\mathcal{B}_W$,  
\begin{equation}
\mathcal{B}_W = \{ -2 < W <  0, 0 < a < \infty, Q=0,   t=0 \},
\label{eq:BangTime}
\end{equation}
where $W$ is defined by a time variable $N$. $N$ is well defined on $\mathcal{B}_W$ and given by (\ref{eq:BasicTimeDE2}). $W(N)$ satisfies the differential equation (\ref{eq:dwdN}) on $\mathcal{B}_W$, where $dW/dN < 0$, $-\infty < N < \infty$, $\lim_{N \rightarrow -\infty} W(N) = 0$,   $\lim_{N \rightarrow +\infty} W(N) = -2$. ($W > -2$)  Also, $\lim_{N \rightarrow -\infty} a(N) = 0$,   $\lim_{N \rightarrow +\infty} a(N) = \infty$.   $a(N)$, in general,  satisfies (\ref{eq:BasicTrans}), implying  $a \propto \e^{N}$. The big bang occurs on the boundary $\mathcal{B}_W$  where $W=0$ and $a = 0$. (See Figure 1)
\medskip
\medskip
\medskip

\noindent
\textbf{Definitions, Assumptions} \; We refer to the set $\mathcal{B}_W$ as the {\em extended big bang state}, since, on it, $W(N)$ varies as a function of $N$, where the interval 
$-\infty <  N < \infty$,  maps into $t=0$.  The variations of $W, a$  on the extended big bang state are, more precisely,  the variations of these two functions for t=0: $a_0, W_0$ as a function of $N$, which can be viewed as a covering, parameterized by $N$, where $a_0=0, W_0=0$.  As defined in the Introduction, the cover of $a_0$ is the set $A$ together with a map, $M_A$ of $A$ onto the point $a_0$, and the cover of $w_0$ is the set $B$ together with a map, $M_B$ of $B$ onto the point $w_0$. $M_A, M_B$ are continuous maps, called covering maps or projections \cite{Munkres:2000}.  On the set  $\mathcal{B}_W$,  $a >0 $ varies in a continuous manner as a function of $N$ as $a \equiv a_0  \propto \e^{N}$.
\medskip\medskip
\medskip\medskip

\noindent
\textbf{Result~3 (Big Bang Mechanism)} \; The big bang can be realized in a mathematically complete manner for $W \geq 0$ with $t >0$  after a flow of the time, $\tilde{N} = -N$, in the extended big bang state,   where $W(\tilde{N}) \equiv W_0(\tilde{N})$ flows from $-2$, corresponding to the cosmological constant, to $0$, and where $\tilde{N}$ varies from $-\infty$ to $+\infty$, respectively. $a(\tilde{N}) \equiv   a_0(\tilde{N})$ varies from $\infty$ to $0$, respectively, where  $a \equiv a_0 \propto \e^{-\tilde{N}}$. Once $a_0 =0$, $W_0 = 0$ in the limit as $\tilde{N} \rightarrow \infty$, then the big bang occurs. Then, for $t >0$, there are infinitely many possible solutions $Q(t), W(t)$, $W \geq 0, Q > 0$, where $a(t) > 0$.  (See Figure \ref{fig:fig2})
\medskip\medskip
\medskip\medskip

\noindent
\textit{Proof of Results~1,2,3.} \; For $t < 0$, consider the time transformation given by
\begin{equation}
\frac{dN}{dt} = - H ,
\label{eq:BasicTimeDE}
\end{equation}
where $H < 0$; or, equivalently,
\begin{equation}
\frac{dt}{dN} = - Q .
\label{eq:BasicTimeDE2}
\end{equation}
Since $H = \dot{a}/a$, equation~(\ref{eq:BasicTimeDE}) implies that the new time variable is simply given by
\begin{equation}
N =  \ln a + \hbox{const} .
\label{eq:BasicTrans}
\end{equation}
\medskip

\noindent
{\bf Remark } \hspace{.05in}It is noted that when $Q=0$, i.e. $|H| = \infty$, the time variable $N$ defined by (\ref{eq:BasicTimeDE2}) is well defined, where $-\infty < N < \infty$, even though $t=0$. The set $C = \{-\infty < N < \infty\}$ represents a cover of $t=0$.  
\medskip

In the new time variable $N$, equations~(\ref{eq:DiffEqusMoreGeneralH}, \ref{eq:DiffEqusMoreGeneralW}, \ref{eq:DiffEqusMoreGeneralOmega}) can be written as the following system of differential equations for $Q$, $W$, and $\Omega$,
\numparts
\begin{eqnarray}
\hspace{-0.5in} \frac{dQ}{dN} = - \case{3}{2} \Big[ (W + w_c + 1) - (W + w_c - w_1) \Omega \Big] Q , \label{eq:dQdN} \\*[4pt]
\hspace{-0.5in} \frac{dW}{dN} = \case{- 3 \sqrt{W + w_c + 1}}{\sqrt{W + w_c + 1} + \sqrt{(w_c + 1) (1 - \Omega)}} \, (W + w_c - 1) \big( W + (1 + w_c) \Omega \big) , \label{eq:dwdN} \\*[4pt]
\hspace{-0.5in} \frac{d\Omega}{dN} = - 3 (W + w_c - w_1) \Omega (1 - \Omega) . \label{eq:dOmegadN}
\end{eqnarray}
\endnumparts
The singularity at $t = 0$ is mapped to a fixed point at $(Q, W, \Omega) = (0, 0, 0)$ to which the solutions flow as $N \to \infty$. The solutions to the original dynamical system can be obtained from the solutions to the transformed differential equations~(\ref{eq:dQdN}, \ref{eq:dwdN}, \ref{eq:dOmegadN}) and the time transformation (\ref{eq:BasicTimeDE2}). Note that, for $t > 0$, the time transformation should be modified by changing the sign in (\ref{eq:BasicTimeDE2}), since $H > 0$ after the big bang. Accordingly, equations~(\ref{eq:dQdN}, \ref{eq:dwdN}, \ref{eq:dOmegadN}) also change sign, and their solutions correspond to trajectories that approach the fixed point $(Q, W, \Omega) = (0, 0, 0)$ as $N \to -\infty$. The solution of the dynamical system $(Q(t), W(t), \Omega(t))$ for $t < 0$ can be branch regularized if it has a unique branch extension to a solution for $t > 0$. The regularized solution is continuous at $t = 0$ where $(Q, W, \Omega) = (0, 0, 0)$. 
\medskip

Setting $\Omega =0, w_c = 1$  we obtain 

\numparts
\begin{eqnarray}
\frac{dQ}{dN} = -\case{3}{2} (W + 2) Q , \label{eq:Qdenew} \\*[4pt]
\frac{dW}{dN} = \case{-3 \sqrt{W + 2}}{\sqrt{W + 2} + \sqrt{2}} \, W^2 . \label{eq:wdenew}
\end{eqnarray}
\endnumparts

This system of differential equations can equivalently be written in terms of the vector $\mathbf{X} \equiv (Q, W)^{\rm T}$ as
\begin{equation}
\frac{d \, \mathbf{X}}{dN} = \mathbf{A} \mathbf{X} + \mathbf{\Delta}(\mathbf{X}) ,
\label{eq:GoodSystem}
\end{equation}
where the matrix $\mathbf{A}$ is given by
\begin{equation}
\mathbf{A} = \left( \begin{array}{cc}
     -3 & 0 \\[4pt]
0 &    0    \end{array} \right) ,
\end{equation}
and the vector field $\mathbf{\Delta}$ is given by
\begin{equation}
\mathbf{\Delta} = \left( \begin{array}{c} -\case{3}{2} Q W \\[4pt]
\frac{-3\sqrt{ (2 + W)}}{ \sqrt{W+ 2}+ \sqrt{2}} \, W^2 \end{array} \right) .
\label{eq:FCase2}
\end{equation}
It is clear that $\mathbf{A} \mathbf{X}$ is the linear order term and $\mathbf{\Delta} = \mathcal{O}(|\mathbf{X}|^2)$.

The matrix  $\mathbf{A}$ is seen to have a zero eigenvalue. This is different from \cite{Xue:2014} where both the eigenvalues of  $\mathbf{A}$ were  nonzero. When both eignenvalues are real and nonzero, then the stable manifold theorem can be used to conclude that the solutions of the linear system yield the solutions of the full nonlinear system by just being slightly perturbed in a smooth manner in a neighborhood of the fixed point at (0,0). However, the existence of a zero eigenvalue gives a degeneracy where the stable manifold theorem cannot be used. In this case, we can allude to the center manifold theorem which says that a slight perturbation of the linear system can yield a complex solution structure in a neighborhood of the fixed point, where there is non-uniqueness of solutions. (see \cite{Guckenheimer:2002}) This can be successfully applied to  (\ref{eq:GoodSystem}). 

In our case, we can explicitly solve  (\ref{eq:GoodSystem}). We obtain from (\ref{eq:Qdenew}),(\ref{eq:wdenew}), 
\begin{equation}
\frac{dQ}{dW} = \frac{Q(W + 2)}{\frac{\sqrt{W+2}}{\sqrt{W+2} + \sqrt{2}} W^2}.
\label{eq:ReducedDE}
\end{equation}

We assume, that prior to the big bang at $t=0$, at a time $t_0 < 0$, $Q(t_0) = Q^0 < 0, W(t_0) = W^0 \geq 0$. The solution is given by,
\begin{equation}
Q(W) = C_0 Q^0 \e^{-\frac{1}{W}(\sqrt{2}\sqrt{W+2} + 2)}  \frac{  \sqrt{1 -      \frac{1}{\sqrt{2}}\sqrt{W+2}    }       }{      \sqrt{1 +  \frac{1}{\sqrt{2}}\sqrt{W +2} }      }  ,
\label{eq:QSolution} 
\end{equation}
where $C_0$ is a positive constant. 
\medskip

\noindent
This implies the following:
\medskip

\noindent
(1.) All trajectories, $Q(W)$, starting in the right half $Q,W$-plane, $W \geq 0$, with initial conditions, $Q^0 < 0, W^0 \geq 0$, corresponding to $t_0 < 0$, approach the fixed point (0,0) corresponding to $t = 0$. More precisely,  
\begin{equation}
\lim_{N \rightarrow \infty} (Q(N), W(N)) = (0,0).
\label{eq:LimitWpositive}
\end{equation}
It is noted that as $N \rightarrow  \infty$, then $t \rightarrow 0$.
\medskip

\noindent
(2.) The only solution emanating from the origin for $W < 0$ is the solution:  $Q = 0, W = W(N)$, where $W(N)$ satisfies the differential equation (\ref{eq:wdenew}), where solutions move to the left on the $W$-axis as $N$  increases ($N > -\infty$). It  is seen from (\ref{eq:ReducedDE}), that another fixed point, in addition to $(Q,W) = (0,0)$, is $(Q,W) = (0, -2)$. This implies 
\begin{equation}
\lim_{N \rightarrow -\infty}  W(N)  = 0     , \hspace{.1in} \lim_{N \rightarrow +\infty} W(N) = -2 .
\label{eq:LimitWnegative}
\end{equation} 
Also, as is seen from (\ref{eq:ReducedDE}), it is necessary that $W > -2$, or equivalently, $w > -1$, satisfying the null energy condition.
\medskip

This solution lies on the set $\mathcal{B}_W$, and on account of the definition of $N$ given by (\ref{eq:BasicTimeDE2}), $t = 0$ on this set. This is because when $Q=0$, this implies $dt = 0$ or equivalently that $t = c =$constant.  However, $t=0$ for $Q=0$ when $W=0$, and therefore for any $W$, $-2 < W < 0$.  The big bang is seen to be extended to $\mathcal{B}_W$ since also, $Q=0$. 
\medskip\medskip

\noindent
(3.) It can be seen by (\ref{eq:BasicTrans}),  that 
\begin{equation}
a \propto \e^{N}.
\end{equation}
Therefore, as $N$ varies on $\mathcal{B}_W$, $a(N) \equiv a_0(N) > 0$, and
\begin{equation}
\lim_{N \rightarrow -\infty}  a(N)  = 0     , \hspace{.1in} \lim_{N \rightarrow +\infty} a(N) = \infty .
\end{equation}
\medskip\medskip

\noindent
(4.) The solutions for $W=0$, constrained to the $Q$-axis, are exponential contractions, as follows from (\ref{eq:Qdenew}). These are give by $Q = Q^0\e^{-3N}$, where $Q \rightarrow 0$ as $N \rightarrow \infty$.   These axis are stable manifolds for the flow.
\medskip\medskip

The description in 1,2,3,4  is summarized in Figure 1, 
which corresponds to  $t \leq 0$.
\medskip
\begin{figure}
\centering
	\includegraphics[width=0.95\textwidth, clip, keepaspectratio]{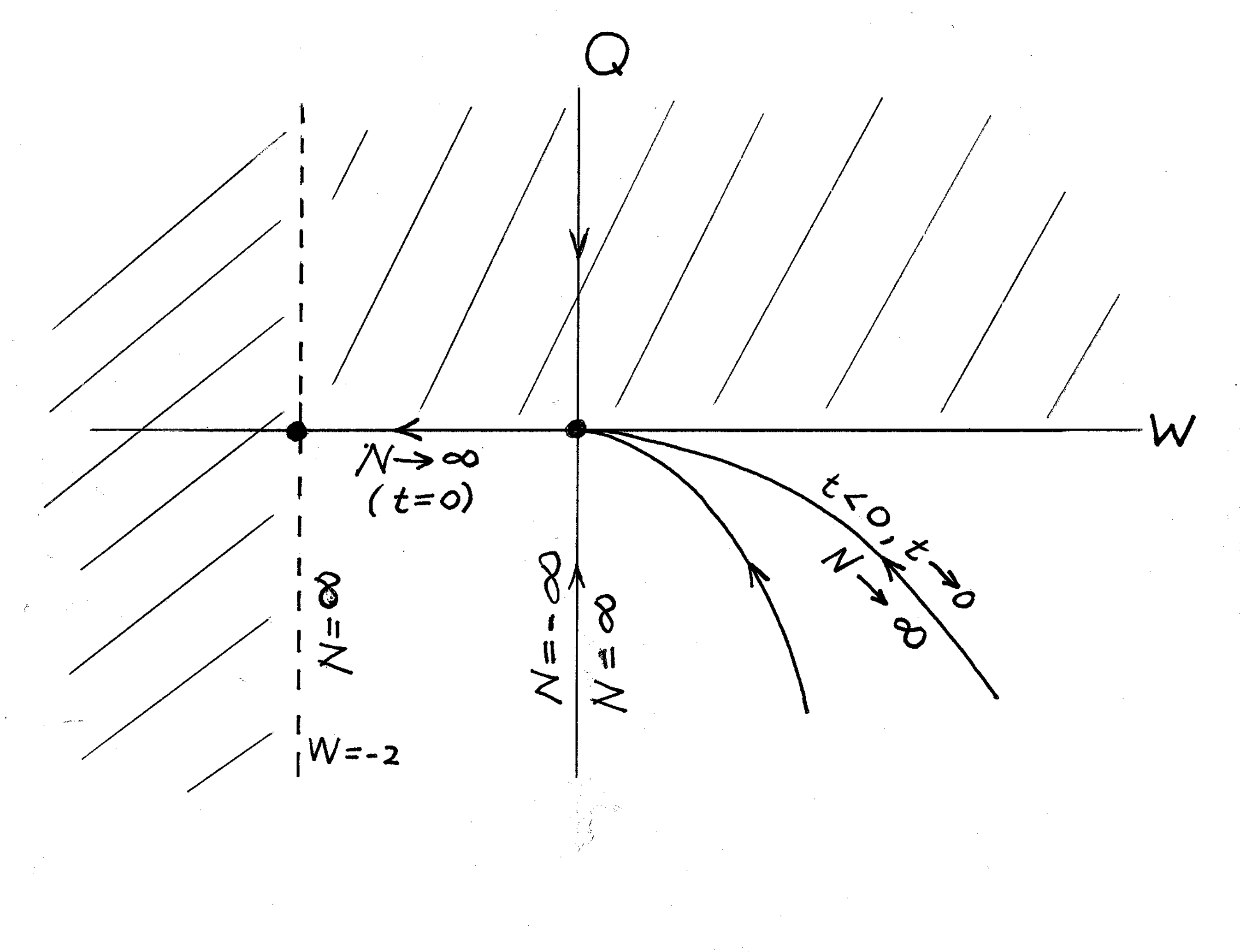}
\label{fig:fig1}
\caption{Contraction, $t < 0$, to the extended big bang state for $t=0$ for $w =1$, then to the big bang state where $w$ decreases to $-1$ with another time variable, $N$.}
\end{figure}

This proves Result 2. 
\medskip

It is remarked that the set of all trajectories leading to the big bang, for $t < 0$, comprise infinitely many branches of the center manifold structure. (see \cite{Guckenheimer:2002}) All these branches are non-unique since they all converge tangentially at the origin. This is different from the stable manifold theorem as in \cite{Xue:2014}, where the manifold branches are finite in number and unique. They are also structurally stable under small perturbation. In the case of the center manifold theorem, the center manifold need not be structurally stable.
\medskip
\medskip\medskip

\noindent
{\em Extending to $t > 0$.}
\medskip\medskip

\noindent
 We make the transformation,
\begin{equation}
 N \rightarrow \tilde{N} = -N
\label{eq:SpecialMap}
\end{equation}
(It is remarked that $\tilde{N}$ is referred to as $N$ in the Introduction and Conclusion for simplicity of notation.)
The system of differential equations, (\ref{eq:Qdenew}), (\ref{eq:wdenew}),  becomes
\numparts
\begin{eqnarray}
\frac{dQ}{d\tilde{N}} = \case{3}{2} (W + 2) Q , \label{eq:Qdenew2} \\*[4pt]
\frac{dW}{d\tilde{N}} = \case{3 \sqrt{W + 2}}{\sqrt{W + 2} + \sqrt{2}} \, W^2 . \label{eq:wdenew2}
\end{eqnarray}
\endnumparts

We also map $t \rightarrow -t$. We now have $t \geq 0$. Also, $ Q \geq  0$. We now start in the left half plane where $W < 0$, after the rest point on the $W$-axis at $W =-2$. Now, we flow to the left of the rest point on the set $\mathcal{B}_W$, where $t=0$, where $ dW/d\tilde{N} > 0$. The following is obtained:
\medskip

\noindent
(i.)  The only solution approaching the big bang at $t=0$, with $Q=0, W=0$, is the solution for $W < 0$ lying in the set $\mathcal{B}_W$, where
\begin{equation}
\lim_{\tilde{N} \rightarrow +\infty}  W(\tilde{N})  = 0 , \hspace{.1in} \lim_{\tilde{N} \rightarrow -\infty}  W(\tilde{N})  = -2 .
\label{eq:LimitWnegative2}
\end{equation} 
\medskip

\noindent
(ii.)   For $W \geq 0$, due to non-uniqueness of solutions emanating from the origin, we can choose the infinitely many solutions for $Q >0$, 
corresponding to $H > 0$ for $t > 0$,  after the bing bang. Since this implies $a(t) > 0$, then the spatial dimensions open up.  These solutions also are defined for $\tilde{N}$ and as $t$ increases, so does $\tilde{N}$.  It is seen that the solutions in the extended big bang and after the big bang are globally defined for all values of $\tilde{N}$ , implying they are mathematically complete in this time variable. 
(see Figure \ref{fig:fig2})
 \medskip
 
 This proves Result 3.
 \medskip
 
 \begin{figure}
\centering
	\includegraphics[width=0.95\textwidth, clip, keepaspectratio]{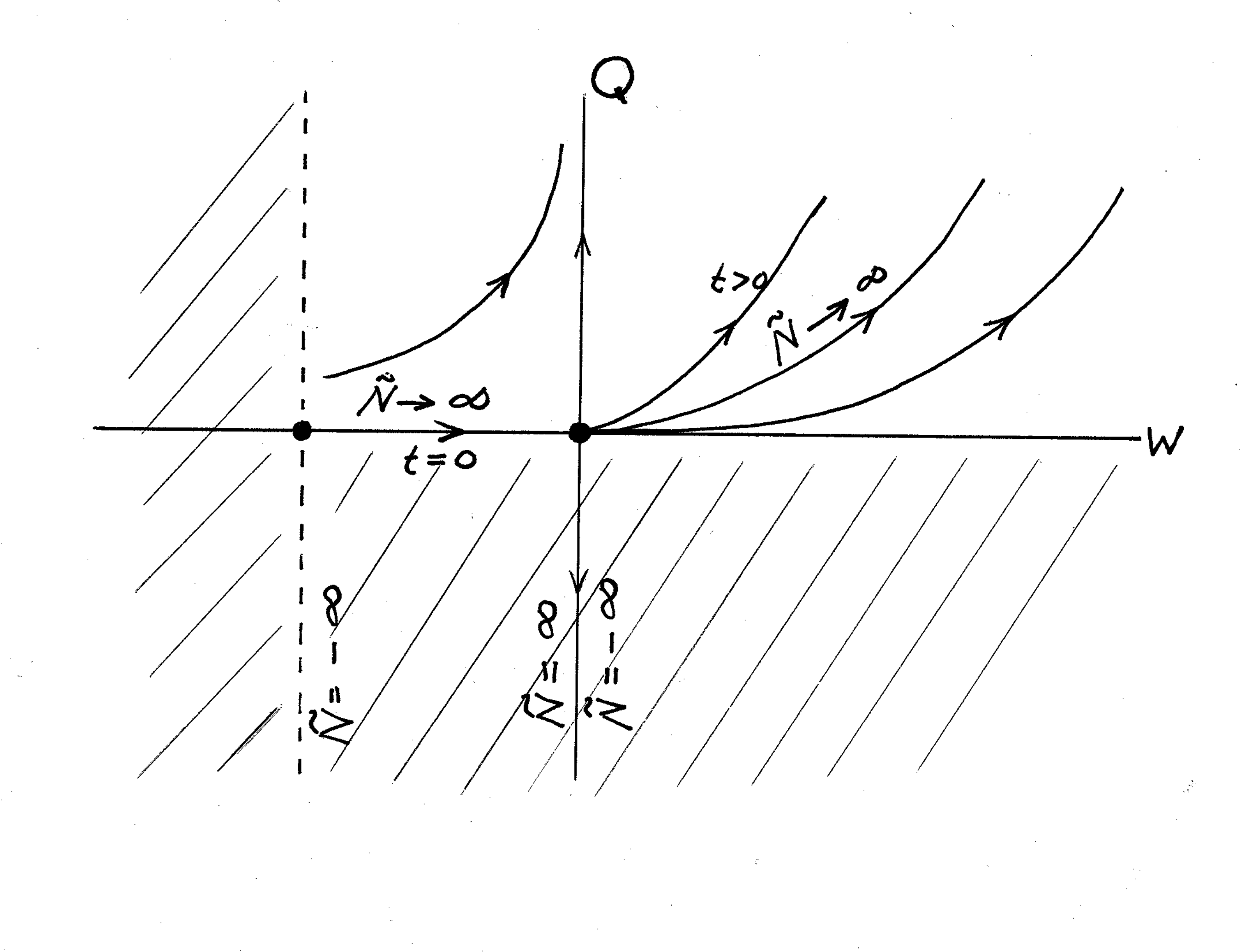}
\caption{Starting in the big bang state for $t=0$ from $w=-1$ to the beginning of the big bang expansion with $w =1$,  using  a time variable, $\tilde{N}$, then  expanding for $t > 0$.}
\label{fig:fig2}
\end{figure}
 \medskip 
 \medskip\medskip
 
As is seen, there is no  unique possible extension from any solution $(W(t), Q(t))$ approaching the origin for $t <0$ with initial condition $(W^0, Q^0), W^0 \geq 0, Q^0 < 0$, at $t = t_0$,  since, after it flows on $\mathcal{B}_W$  to the left end point $W = -2$  (see Figure 1), and then flows from $W = -2$ on $\mathcal{B}_W$ to the right end point $W =0$, there is no unique extension to $W \geq 0$ for $ t > 0$ (see Figure \ref{fig:fig2}).   
 \medskip

This proves Result 1.
\medskip\medskip
\medskip

\noindent
{\bf Remark 1} \hspace{.05in} In the proof that there are infinitely many extensions at $t=0$, we used the extended big bang state to describe this.   This was not necessary as it could have been proven by skipping the flow on $\mathcal{B}_W$, and just have gone from a solution approaching the big bang at $t=0$, for $t <0$, to $a=0$, $W=0$ at $t=0$, then directly connect to $t > 0$.  The use of $\mathcal{B}_W$ in our proof was done as an alternative and for a little more generality.
\medskip

\noindent 
{\bf Remark 2} \hspace{.05in}  It is noted that the definition of the variable $N$ by (\ref{eq:BasicTimeDE}) was used in \cite{Xue:2014}. In that paper an extended big bang type set was not was not defined.  Instead all solution matchings were done at $Q=1/H =0$ and $W=w-wc = 0$, $w_c > 1$. 
Also, in \cite{Xue:2014}(see page 10, (34a, 34b)) there is a rest point for the relevant system of differential equations at $W = 1-w_c < 0$(where for this flow we require $1 <w_c < 2$ to satisfy the null energy condition). However, this is a rest point only for the flow restricted to the $W$-axis for $Q=0$ and not for the full $W,Q$-flow.  This is because the relevant  differential equation in \cite{Xue:2014}(equation 34a) for $Q$ is given by $dQ/dN = (-3/2)(W+w_c+1)Q$. This is unlike  (\ref{eq:Qdenew}), (\ref{eq:wdenew}). In this system, since the point $(-2,0)$ is a rest point for the full flow in both variables $(W,Q)$, this yields directions to approach $(-2, 0)$ off of the $W$-axis.  In this sense, it is a dynamically more significant rest point. 
\medskip

\noindent
{\bf Remark 3} \hspace{.05in} It is remarked that the situation in \cite{Xue:2014} for $w_c > 1$ is entirely different. In that case, for each solution  $(W(t), Q(t))$  approaching the origin with a given initial condition at $t=t_0<0$, there is a unique extension to $t > 0$ through the big bang by appealing to the stable manifold theorem, provided $w_c$ satisfies relative prime number conditions.  The big bang singularity in that case can be regularized. 
In that paper, the solution extensions were patched at $W = w-w_c = 0$.  
\medskip\medskip\medskip\medskip
 
\subsection{Different Big Bang Starting Scenarios and Interpretations }
\label{subsec:differentmechanisms}
 \medskip
 
 \noindent
 The big bang can be viewed as starting by two different ways, based on Results 2,3.  
 \medskip
 
\noindent 
{\em Scenario A}:   we start from a contracting universe to the extended big bang then to an expanding universe. More precisely, we start in Result 2 and shown in Figure 1, where one starts at a time $t_0 < 0$ in the contracting universe at $W^0, Q^0$. The solution $(W(t),Q(t))$ approaches $(W_0, Q_0) \equiv (0,0)$ as $t \rightarrow 0^-$, or as $N \rightarrow \infty$. The solution then enters the extended big bang state at $t=0$, where $Q=0$. It moves from $W (\equiv W_0)$ to $W=-2$ as $N$ varies between $-\infty$ and $\infty$, and $a \equiv a_0$ varies from $\infty$ to $0$.  One then connects the point $(W,Q) = (-2, 0)$ in Figure 1 with the same point in Figure \ref{fig:fig2}. The direction of the flow reverses, and, by Result 2,  the time variable $\tilde{N} = -N$ in the set $\mathcal{B}_W$ varies from $-\infty$ to $\infty$ where $a \rightarrow 0, W \rightarrow 0$. Then the big bang occurs and the universe expands for $t >0$, or where $\tilde{N}$ increases from $-\infty$ with infinitely many possible extensions $(W(t), Q(t)$. 
\medskip\medskip

\noindent
{\em Scenario B}: The second possible starting point would just be the situation described at the end of Scenario A  shown in by Figure \ref{fig:fig2},  where one just starts within $\mathcal{B}_W$ where $W \equiv W_0$ varies from $-2$ to $0$ and $a \equiv a_0$ varies from $\infty$ to $0$, as $\tilde{N}$ varies from $-\infty$ to $\infty$. As mentioned in the Introduction, this leads to possible physical interpretation of how the big bang could start.  One can imagine an expanding universe that expands until all the matter and energy dissipates, after a sufficiently long time, say trillions of years where $a$ is extremely large, say $a \approx \infty$ ($a < \infty$). Also, at this epoch, it is reasonable to assume that $w \approx -1$ ($w$ slightly less than $-1$).   Since all matter and energy have dissipated, then there is no way to distinguish one location from another.  All locations are equivalent.  In that sense, the universe can be viewed as being equivalent to a single point, which would serve as a starting point for the big bang, where time is reset to $t=0$. The initial point for $t=0$ would be near the left endpoint of $\mathcal{B}_W$ , where $a \approx \infty$ and $w \approx -1$. The time variable $\tilde{N} \approx -\infty$ ($\tilde{N} < \infty$). The variable $\tilde{N}$ increases from $-\infty$ where $a(\tilde{N})$ decreases and $w$ increases.  As $\tilde{N} \rightarrow \infty$, then $a(\tilde{N}) \rightarrow 0$ and $w \rightarrow 1$. In the limit, $a=0$, $w=1$, which is the right endpoint of $\mathcal{B}_W$ .  The big bang singularity is reached at $t=0$ where expansion starts for increasing $t > 0$. 

Scenerio B is similar to what is described in the short story entitled {\em The Last Question} , by I. Asimov \cite{Asimov:1956}.  This supports the model of the big bang he described in that story. 
 \medskip

\section{Generalization to Additional Energy Components}
\label{sec:3}
 \medskip

\noindent
The previous results are generalized by including other energy components in addition to the scalar field. As described in Section \ref{sec:1}, we let $\Omega_m$ be a density parameter corresponding to an energy density $\rho_m$, with a constant equation of state $w_m$, where $w_m \in \{ 0, \frac{1}{3}, -1, -\frac{1}{3}, 1 \}$. As required below, we exclude $w_m = 1$.  Let $w_1$ represent one of these, with corresponding density parameter $\Omega_1$, we label more simply as $\Omega$. The system of differential equations (\ref{eq:dQdN}), (\ref{eq:dwdN}), (\ref{eq:dOmegadN}) are considered with $w_c=1$,  
\medskip
\numparts
\begin{eqnarray}
\hspace{-0.5in} \frac{dQ}{dN} = -3Q  - \case{3}{2} Q \Big[ (W + 2) - (W + 1 - w_1) \Omega \Big]  , \label{eq:dQdN2} \\*[4pt]
\hspace{-0.5in} \frac{dW}{dN} = \case{- 3 \sqrt{W + 2}}{\sqrt{W + 2} + \sqrt{2 (1 - \Omega)}} \, W \big( W + 2 \Omega \big) , \label{eq:dwdN2} \\*[4pt]
\hspace{-0.5in} \frac{d\Omega}{dN} = - 3 (1 - w_1) \Omega    -3W\Omega + 3(1-w_1)\Omega^2 + 3W\Omega^2 \label{eq:dOmegadN2} .
\end{eqnarray}
\endnumparts
There are rest points at $(Q, W, \Omega) = (0,0,0), (0,-2,0)$.  

Results 1,2,3 generalize in a natural way, where now an additional dimension is added due to $\Omega$, where $\Omega \rightarrow 0$ as $t \rightarrow 0$. The set $\mathcal{B}_W$ is generalized to
\begin{equation}
\mathcal{B}_W = \{ -2 < W <  0, Q=0, \Omega = 0, 0 < a < \infty,  t=0 \}.
\label{eq:BangTime2}
\end{equation}
The flow of the trajectories, $Q(N), W(N), \Omega(N)$ is obtained by analyzing   (\ref{eq:dQdN2}), (\ref{eq:dwdN2}), (\ref{eq:dOmegadN2}).  This system is valid for $t \leq  0$, where it is assumed that $Q \leq 0$. As was done previously in Section \ref{sec:2}, we consider an initial condition,  $Q^0, W^0, \Omega^0$, corresponding to $t = t_0 < 0$, where  $Q^0 < 0$, $W^0 > 0$.  The corresponding trajectory for $Q,W,\Omega$ approaches $(0,0,0)$ as $ t \rightarrow 0$, or equivalently as $N \rightarrow +\infty$. There are infinitely many trajectories that approach the origin in the lower half space corresponding to $(Q,W,\Omega)$, $Q < 0$, $W > 0$. The only solution that exists as an extension to $W < 0$  at the origin from any of these trajectories, lies on the set $\mathcal{B}_W$ on the negative $W$-axis. The flow on $\mathcal{B}_W$ is given from the system  (\ref{eq:dQdN2}), (\ref{eq:dwdN2}), (\ref{eq:dOmegadN2}) by setting $Q=\Omega = 0$ yielding (\ref{eq:wdenew}) as in the previous section, where $dW/dN < 0$, where the trajectory, $(0, W(N), 0)$  has the limits, 
\begin{equation}
\lim_{N \rightarrow +\infty} (0,W(N),0) = (0, -2, 0), \hspace{.1in}  \lim_{N \rightarrow -\infty} (0,W(N),0) = (0, 0, 0).
\label{eq:Limits}
\end{equation}
\medskip\medskip

Of particular importance for this paper is what happens when we look at the flow in the other direction, via the map, $N \rightarrow -N = \tilde{N}$, and $t \rightarrow -t$, where now, using the same symbol for time, we have $t \geq 0$. Also, $Q \geq 0$.  The differential equations  (\ref{eq:dQdN2}), (\ref{eq:dwdN2}), (\ref{eq:dOmegadN2}) become, 
\numparts
\begin{eqnarray}
\hspace{-0.5in} \frac{dQ}{d\tilde{N}} = 3Q  + \case{3}{2} Q \Big[ (W + 2) - (W + 1 - w_1) \Omega \Big]  , \label{eq:dQdN3} \\*[4pt]
\hspace{-0.5in} \frac{dW}{d\tilde{N}} = \case{ 3 \sqrt{W + 2}}{\sqrt{W + 2} + \sqrt{2 (1 - \Omega)}} \, W \big( W + 2 \Omega \big) , \label{eq:dwdN3} \\*[4pt]
\hspace{-0.5in} \frac{d\Omega}{d\tilde{N}} = 3 (1 - w_1) \Omega    +3W\Omega - 3(1-w_1)\Omega^2 - 3W\Omega^2 \label{eq:dOmegadN3} .
\end{eqnarray}
\endnumparts

         We now begin on the set $\mathcal{B}_W$  at the rest point $(0, -2, 0)$, corresponding to $\tilde{N} = -\infty$. The system of differential equations   (\ref{eq:dQdN3}), (\ref{eq:dwdN3}), (\ref{eq:dOmegadN3})  reduces to the differential equation (\ref{eq:wdenew2})  for $W(\tilde{N})$ where $W$ goes from the rest point at $(0,-2,0)$ to the rest point at the origin and 
\begin{equation}
\lim_{\tilde{N} \rightarrow -\infty} (0,W(\tilde{N}),0) = (0, -2, 0) \hspace{.1in}  \lim_{\tilde{N} \rightarrow +\infty} (0,W(\tilde{N}),0) = (0, 0, 0).
\label{eq:Limits2}
\end{equation}
The origin is the right endpoint of $\mathcal{B}_W$ and is the big bang singularity. When $t >0$ the curves $Q(\tilde{N}), W(\tilde{N}), \Omega(\tilde{N})$  emanate from the origin after the big bang as $\tilde{N}$ increases, or, equivalently, as $t$ increases,  and as $\tilde{N}$ decreases, we see that
\begin{equation}
\lim_{\tilde{N} \rightarrow -\infty} (Q(\tilde{N}), W(\tilde{N}), \Omega(\tilde{N})) = (0, 0, 0).
\label{eq:Limits3}
\end{equation}
The behavior of these curves and the flow, in general, is determined by examining the solutions of (\ref{eq:dQdN2}), (\ref{eq:dwdN2}), (\ref{eq:dOmegadN2}) on different subspaces.

As in the planar case considered previously shown in Figure \ref{fig:fig2}, we require $W > -2$. Also, since $t \geq 0$, we require $ Q \geq 0$. In what follows, it is required that $|\Omega|$ is sufficiently small. This yields the domain, $D$, for the variables,
\begin{equation}
D = \{ (Q,W,\Omega) |  W > -2, Q \geq 0, |\Omega| < \delta \},
\label{eq:Domain}
\end{equation}
where $\delta$ is sufficiently small. 

\begin{figure}
\centering
	\includegraphics[width=1.2\textwidth, clip, keepaspectratio]{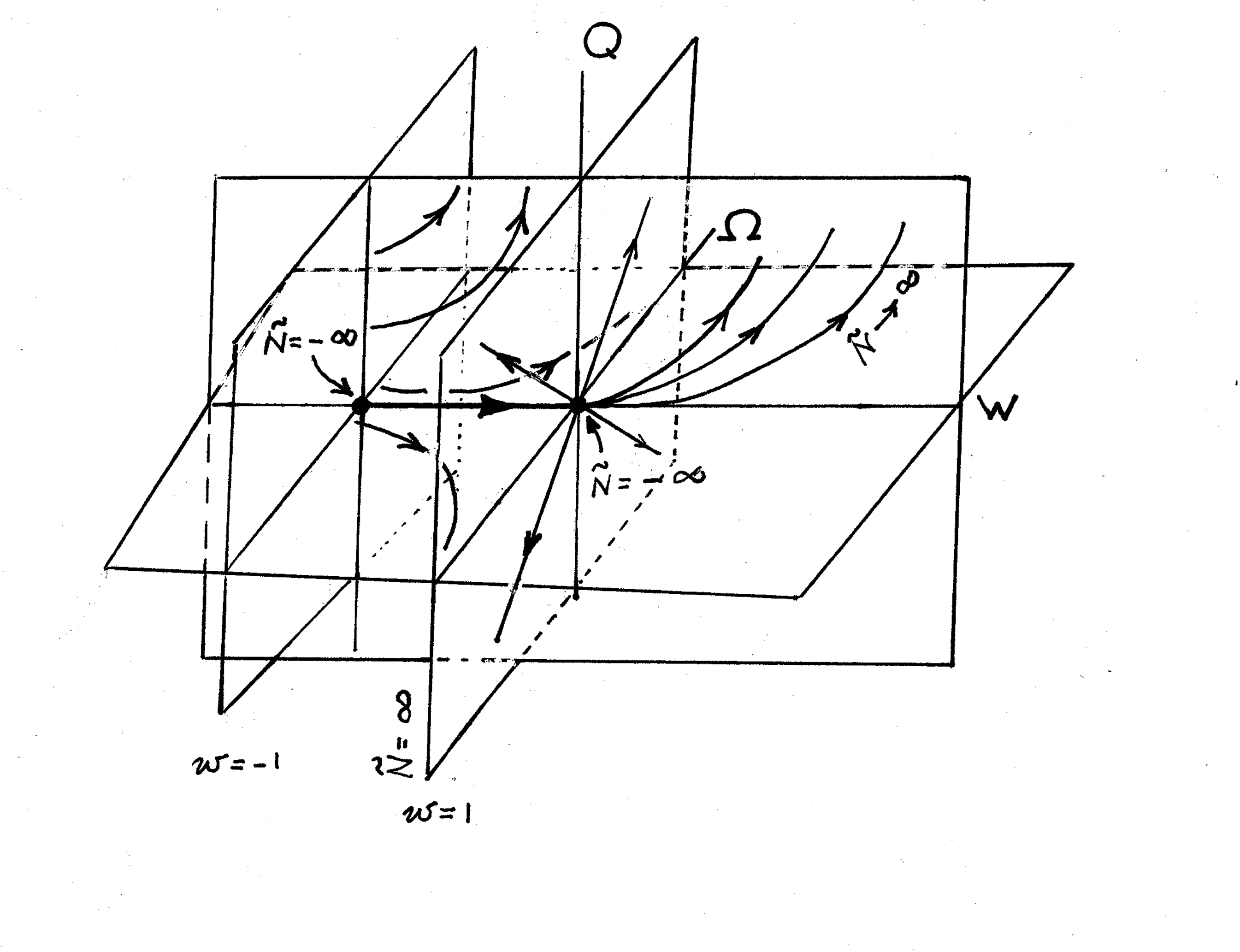}
\caption{Addition of another energy component $\Omega$ to Figure \ref{fig:fig2}.  }
\label{fig:fig3}
\end{figure}

On the subspace $\Omega = 0$, the system of differential equations reduces to the invariant subsystem (\ref{eq:Qdenew2}), (\ref{eq:wdenew2}) considered in the previous section, where the flow is shown in Figure \ref{fig:fig2}. Also, see Figure \ref{fig:fig3}.

On the subspace $W=0$, the system of differential equations reduces to the invariant subsystem, 
\begin{eqnarray}
\hspace{-0.5in} \frac{dQ}{d\tilde{N}} = 3Q + \case{3}{2} Q \Big[  (1 - w_1) \Omega \Big]    , \label{eq:dQdN3Prime} \\*[4pt]
\hspace{-0.5in} \frac{d\Omega}{d\tilde{N}} = 3 (1 - w_1)(\Omega   -  \Omega^2)  \label{eq:dOmegadN3Prime}.
\end{eqnarray}
The behavior of the solutions for these differential equations for $|\Omega|$ sufficiently near the origin is seen to be dominated by the first order linear system,
\begin{eqnarray}
\hspace{-0.5in} \frac{dQ}{d\tilde{N}} = 3Q    , \label{eq:dQdN3Prime2} \\*[4pt]
\hspace{-0.5in} \frac{d\Omega}{d\tilde{N}} = 3 (1 - w_1)\Omega  \label{eq:dOmegadN3Prime2}.
\end{eqnarray}
The solutions of (\ref{eq:dQdN3Prime2}), (\ref{eq:dOmegadN3Prime2}) are given by
\begin{equation}
Q(\tilde{N}) \propto \e^{3\tilde{N}},   \hspace{.2in}  \Omega(\tilde{N})  \propto \e^{3(1-w_1)N}
\label{eq:BasicSolutions}
\end{equation}
These solutions represent exponential expansions, since it is assumed $w_1 < 1$.
The two solutions $X_1 = (Q(\tilde{N}), 0), X_2 =(0, \Omega(\tilde{N}))$ form a basis for all solutions on the subspace $W=0$ for(\ref{eq:dQdN3Prime2}), (\ref{eq:dOmegadN3Prime2}).  

By the stable manifold theorem \cite{Guckenheimer:2002}, the solutions for  (\ref{eq:dQdN3Prime}), (\ref{eq:dOmegadN3Prime}) are very close to those of (\ref{eq:dQdN3Prime2}), (\ref{eq:dOmegadN3Prime2}) near the origin, and can be mapped 1-1 and continuously onto them, by a mapping $h(Q,W)$.  (The stable manifold theorem can be applied since the higher order terms in  (\ref{eq:dQdN3Prime}), (\ref{eq:dOmegadN3Prime}) are quadratic in $Q, \Omega$.) This implies that the basis solutions $ X_1, X_2$ , can be used as a basis for the solutions of  (\ref{eq:dQdN3Prime}), (\ref{eq:dOmegadN3Prime}) on $\Omega=0$ as $Y_1= h(X_1),Y_2 =  h(X_2)$.   Since the actual solutions to   (\ref{eq:dQdN3Prime}), (\ref{eq:dOmegadN3Prime}) will lie very close to those of  (\ref{eq:dQdN3Prime2}), (\ref{eq:dOmegadN3Prime2}), $Y_1(\tilde{N}), Y_2(\tilde{N})$ will lie very close to the $Q$-axis and $\Omega$-axis, respectively. 

The solutions generated by the basic solutions,  (\ref{eq:BasicSolutions}), as linear combinations form expansions on $W=0$. This is illustrated in Figure \ref{fig:fig3}. 

On the subspace $Q = 0$, the system of differential equations reduces to the invariant subsystem,
\numparts
\begin{eqnarray}
\hspace{-0.5in} \frac{dW}{d\tilde{N}} = \case{ 3 \sqrt{W + 2}}{\sqrt{W + 2} + \sqrt{2(1 - \Omega}} \, W(W + 2 \Omega), \label{eq:dwdN3Prime3} \\*[4pt]
\hspace{-0.5in} \frac{d\Omega}{d\tilde{N}} = 3 (1 - w_1) \Omega    +3W\Omega - 3(1-w_1)\Omega^2 - 3W\Omega^2 \label{eq:dOmegadN3Prime3} .
\end{eqnarray}
\endnumparts
For $W < 0$, and for $\Omega$ sufficiently near $0$,  it is seen that no solution goes to origin except one for $\Omega=0$ on the extended  big bang set. All other solutions have initial conditions of the form $(W^0, \Omega^0) = (-2, \Omega^0)$ whose trajectories are all transversal to the line $W = -2$ and initially move to the right since $dW/d\tilde{N} > 0$, then as $\tilde{N} \rightarrow \infty$ become asymptotic to the $\Omega $-axis and moving away from the origin as is shown in Figure \ref{fig:fig3}. For $W > 0$, no solutions emanate from the origin, since for $t >0$, it is necessary that $Q > 0$.
 
 When the solutions move off of the planes, $Q=0, W=0, \Omega=0$, then the resulting flow of the solutions can be obtained near the origin as a natural composition of the solutions on the planes, shown in Figure \ref{fig:fig3}. It is necessary to assume that $|\Omega|$ is sufficiently small as follows from our analysis.  As is seen, only the extended big bang solution is able to reach the big bang. All solutions are parameterized by $\tilde{N}$, where the limiting values of $\tilde{N}$ are indicated in Figure \ref{fig:fig3}. 
 
\section{Conclusion} \label{sec:4}

\noindent
It is shown that the big bang singularity cannot be regularized in the case of a time varying equation of state where $w \rightarrow 1$ as the universe contracts for $ t \rightarrow 0$. This is because there are infinitely many possible extensions of $w, a, \Omega$, all equally likely. This nonuniquenss yields infinitely many possible expanding universes in this case.  This is modeled using a standard set of Friedmann equations for $H$ and differential equations for $w$, $\Omega$. This is different from the situation where $w \rightarrow w_c >1$, as $ t \rightarrow 0$ that yields a unique extension for discrete values of $w_c$ \cite{Xue:2014}. Thus, the case $w=1$ yields a much more robust big bang. 

A dynamical mechanism is given for the big bang by showing that at $t=0$, it can be extended to a set, $\mathcal{B}_W$, which introduces a new time variable $-\infty < N < \infty$, where both $w \equiv w_0$ and $a \equiv a_0$ vary as a function of N, and where $|H| = \infty$. When $N \rightarrow \infty$, then $a(N) \rightarrow 0, w(N) \rightarrow 1$ and the big bang occurs. The dynamic flow for the big bang is mathematically complete as a function of $N$. One of the interpretations of this is related to a popular short story by I. Asimov \cite{Asimov:1956}.  The extended big bang state and resulting dynamics is generalized for other energy components. 

It is interesting to note that the nonuniqueness of extensions may be related to the nonuniqueness of extension through an anti-gravity loop studied by Bars, et. al. in \cite{Bars:2011aa}. This would be interesting to study for future work.

The methods presented in this paper are purely mathematical, using dynamical systems.

\ack

I would like to thank David Spergel, Paul Steinhardt, and Frans Pretorius for helpful discussions which help guide this work.

\end{document}